\documentclass[twocolumn,tighten]{aastex63}

\usepackage{graphicx}
\usepackage{amsmath}
\usepackage{apjfonts}
\usepackage{scalerel}
\usepackage{tikz}
\usetikzlibrary{svg.path}

\definecolor{orcidlogocol}{HTML}{A6CE39}
\tikzset{
  orcidlogo/.pic={
    \fill[orcidlogocol] svg{M256,128c0,70.7-57.3,128-128,128C57.3,256,0,198.7,0,128C0,57.3,57.3,0,128,0C198.7,0,256,57.3,256,128z};
    \fill[white] svg{M86.3,186.2H70.9V79.1h15.4v48.4V186.2z}
                 svg{M108.9,79.1h41.6c39.6,0,57,28.3,57,53.6c0,27.5-21.5,53.6-56.8,53.6h-41.8V79.1z M124.3,172.4h24.5c34.9,0,42.9-26.5,42.9-39.7c0-21.5-13.7-39.7-43.7-39.7h-23.7V172.4z}
                 svg{M88.7,56.8c0,5.5-4.5,10.1-10.1,10.1c-5.6,0-10.1-4.6-10.1-10.1c0-5.6,4.5-10.1,10.1-10.1C84.2,46.7,88.7,51.3,88.7,56.8z};
  }
}

\newcommand\orcidicon[1]{\href{https://orcid.org/#1}{\mbox{\scalerel*{
\begin{tikzpicture}[yscale=-1,transform shape]
\pic{orcidlogo};
\end{tikzpicture}
}{|}}}}

\usepackage{comment}
\usepackage{array,multirow,color,tablefootnote}
\usepackage{xcolor}
\begin{document}
\title{TESS Revisits WASP-12: Updated Orbital Decay Rate and Constraints on Atmospheric Variability} 

\author[0000-0001-9665-8429]{Ian~Wong \orcidicon{0000-0001-9665-8429}}
\altaffiliation{NASA Postdoctoral Program Fellow}
\affiliation{NASA Goddard Space Flight Center, 8800 Greenbelt Road, Greenbelt, MD 20771, USA; \href{mailto:ian.wong@nasa.gov}{ian.wong@nasa.gov}}

\author[0000-0002-1836-3120]{Avi~Shporer \orcidicon{0000-0002-1836-3120}}
\affil{Department of Physics and Kavli Institute for Astrophysics and Space Research, Massachusetts Institute of Technology, Cambridge, MA 02139, USA}

\author[0000-0003-2527-1475]{Shreyas~Vissapragada \orcidicon{0000-0003-2527-1475}}
\affiliation{Division of Geological and Planetary Sciences, California Institute of Technology, 1200 East California Boulevard, Pasadena, CA 91125, USA}

\author[0000-0002-0371-1647]{Michael~Greklek-McKeon \orcidicon{0000-0002-0371-1647}}
\affiliation{Division of Geological and Planetary Sciences, California Institute of Technology, 1200 East California Boulevard, Pasadena, CA 91125, USA}

\author[0000-0002-5375-4725]{Heather~A.~Knutson \orcidicon{0000-0002-5375-4725}}
\affiliation{Division of Geological and Planetary Sciences, California Institute of Technology, 1200 East California Boulevard, Pasadena, CA 91125, USA}

\author[0000-0002-4265-047X]{Joshua~N.~Winn
\orcidicon{0000-0002-4265-047X}}
\affiliation{Department of Astrophysical Sciences, Princeton University, Princeton, NJ 08544, USA}

\author[0000-0001-5578-1498]{Bj{\" o}rn~Benneke \orcidicon{0000-0001-5578-1498}}
\affiliation{Department of Physics and Institute for Research on Exoplanets, Universit{\' e} de Montr{\' e}al, Montr{\' e}al, QC, Canada}

\begin{abstract}
After observing WASP-12 in the second year of the primary mission, the Transiting Exoplanet Survey Satellite (TESS) revisited the system in late 2021 during its extended mission. In this paper, we incorporate the new TESS photometry into a reanalysis of the transits, secondary eclipses, and phase curve. We also present a new $K_s$-band occultation observation of WASP-12b obtained with the Palomar/Wide-field Infrared Camera instrument. The latest TESS photometry spans three consecutive months, quadrupling the total length of the TESS WASP-12 light curve and extending the overall time baseline by almost two years. Based on the full set of available transit and occultation timings, we find that the orbital period is shrinking at a rate of $-29.81 \pm 0.94$ ms yr$^{-1}$. The additional data also increase the measurement precision of the transit depth, orbital parameters, and phase-curve amplitudes. We obtain a secondary eclipse depth of $466 \pm 35$ ppm, a $2\sigma$ upper limit on the nightside brightness of 70 ppm, and a marginal $6\overset{\circ}{.}2 \pm 2\overset{\circ}{.}8$ eastward offset between the dayside hotspot and the substellar point. The voluminous TESS dataset allows us to assess the level of atmospheric variability on timescales of days, months, and years.  We do not detect any statistically significant modulations in the secondary eclipse depth or day--night brightness contrast. Likewise, our measured $K_s$-band occultation depth of $2810 \pm 390$ ppm is consistent with most $\sim$2.2 $\mu$m observations in the literature.
\end{abstract}


\section{Introduction}\label{sec:intro}

Since its discovery in 2009 \citep{wasp12}, WASP-12b has become one of the most intensively studied exoplanets. This archetypal ultra-hot Jupiter, with a mass of 1.5 $M_{\rm Jup}$ and an inflated radius of 1.9 $R_{\rm Jup}$ \citep{collins2017}, has a 1.09 day orbit around a 6150\,K F-type star \citep[e.g.,][]{stassun2019}. The intense stellar irradiation and the high dayside equilibrium temperature of $\approx$2600 K have made WASP-12b a prime target for atmospheric study. Extensive spectroscopic and photometric transit and secondary eclipse measurements have been obtained at infrared wavelengths \citep{campo2011,croll2011,cowan2012,crossfield2012,zhao2012,sing2013,stevenson2014a,stevenson2014b,croll2015,kreidberg2015,bell2019,arcangeli2021} and visible wavelengths \citep{lopez2010,copperwheat2013,fohring2013,sing2013,bell2017,hooton2019,vonessen2019}. Some highlights from these observations include the detection of water vapor in the upper atmosphere along the day--night terminator \citep{kreidberg2015} and the very low dayside reflectivity \citep{bell2017}. In addition, full-orbit infrared phase curves from the Spitzer Space Telescope have provided a glimpse of the planet's global temperature distribution \citep{cowan2012,bell2019}.

WASP-12b is notable in several respects. First, the planet is undergoing atmospheric mass loss. Repeated observations of the planet's transit in the near-ultraviolet and in H-alpha have revealed a strongly asymmetric signal with an early ingress, suggestive of a distended circumplanetary disk of stripped material \citep{fossati2010,haswell2012,nichols2015,jensen2018}. This scenario is supported by the recent analysis of the Spitzer 4.5 $\mu$m phase curve, which displays a significant modulation with a period equal to half of the orbital period that could be explained by CO emission from escaping gas flowing from the planet into the host star \citep{bell2019}.

Another peculiarity of WASP-12b is the apparent variation in its dayside atmospheric brightness. Mutually discrepant secondary eclipse depths have been reported at various optical wavelengths \citep{lopez2010,fohring2013,hooton2019,vonessen2019}, as well as in the 2 $\mu$m region \citep{crossfield2012,zhao2012,croll2015}. In light of other purported detections of atmospheric variability in hot Jupiters (e.g., \citealt{armstrong2016,jackson2019}; but also see \citealt{lally2020}), the question of whether these differences are real or are instead caused by instrumental effects and/or observing conditions has come to the fore. Meanwhile, recent theoretical work on the atmospheric dynamics of hot Jupiters has shown that hydrodynamic instabilities and other transient atmospheric processes may lead to orbit-to-orbit evolution in the global average temperature, the effects of which may be detectable with current and near-future instruments \citep[e.g.,][]{tan2019,komacek2020,tan2020}.

Lastly, WASP-12b is the only exoplanet for which there is clear evidence for a gradually shrinking orbital period. Evidence for a change in period was first reported by \citet{m16}. Subsequent transit and occultation observations solidified the detection and ruled out apsidal precession and the R{\o}mer effect as explanations for the changing period \citep{patra2017,yee2020}. The best explanation appears to be tidal orbital decay, which has motivated theorists to try to understand the observed rate of decay in terms of tidal dissipation mechanisms within the star \citep[e.g.,][]{weinberg2017,bailey2019,ma2021}

From 2019 December to 2020 January, the Transiting Exoplanet Survey Satellite (TESS; \citealt{ricker2015}) observed WASP-12 as part of its ongoing mission to search the entire sky for transiting exoplanets. The light curves provided additional transit timings, which were used to refine the estimate of the planet's orbital decay rate \citep{turner2021}. Meanwhile, the planetary phase curve was analyzed by \citet{owens2021} and \citet{wong2021year2}, yielding a robust secondary eclipse measurement and broad constraints on the day--night temperature contrast. No evidence for significant orbit-to-orbit variability in the dayside brightness was found in the TESS photometry. TESS has since revisited the WASP-12 system during its ongoing extended mission, providing an additional two and a half months of nearly continuous photometric observations.

In this paper, we present a follow-up study of WASP-12b's orbital decay, phase curve, and potential atmospheric variability, incorporating the latest TESS light curves from 2021 as well as a new ground-based $K_s$-band observation of an occultation.  The relevant datasets are described in Section~\ref{sec:obs}. An overview of our light-curve modeling techniques and our results is provided in Section~\ref{sec:ana}. In Section~\ref{sec:dis}, we use our measurements to update the orbital ephemeris and derive constraints on the level of atmospheric variability. Finally, the main findings of our work are summarized in Section~\ref{sec:conclusion}.

\section{Observations}\label{sec:obs}

\subsection{TESS Light Curves}\label{subsec:tess}

\begin{deluxetable*}{ccccccc}[t]
\tablewidth{0pc}
\setlength{\tabcolsep}{14pt}
\renewcommand{\arraystretch}{0.9}
\tabletypesize{\footnotesize}
\tablecaption{
    List of TESS Light-curve Segments
    \label{tab:segments}
}
\tablehead{
    \colhead{Segment\tablenotemark{\scriptsize$\mathrm{a}$}}                     &
    \colhead{$n_{\mathrm{raw}}$\tablenotemark{\scriptsize$\mathrm{b}$}}                     &
    \colhead{$n_{\mathrm{trimmed}}$\tablenotemark{\scriptsize$\mathrm{b}$}} &
    \colhead{$T_{\mathrm{start}}$\tablenotemark{\scriptsize$\mathrm{c}$}}  &
    \colhead{$T_{\mathrm{end}}$\tablenotemark{\scriptsize$\mathrm{c}$}} & 
    \colhead{Order\tablenotemark{\scriptsize$\mathrm{d}$}} & 
    \colhead{Comments}
}
\startdata
20-1-1 & 3910 & 3101 &  843.510 &  847.922 & 0 & Removed 1.00 day from start\\
20-1-2 & 3870 & 3788 &  847.935 &  853.298 & 0 & \\
20-1-3 & 1119 & 1094 &  853.310 &  854.862 & 1 & \\
20-2-1 & 3029 & 2640 &  858.200 &  861.944 & 1 & Removed 0.25 day from start\\
20-2-2 & 3959 & 3874 &  861.955 &  867.444 & 1 & \\
20-2-3 & 989 & 952 &  867.455 &  868.827 & 1 & \\
43-1-1 & 2922 & 2870 & 1474.172 & 1478.208 & 0 & \\
43-1-2 & 2880 & 2833 & 1478.219 & 1482.209 & 1 & \\
43-1-3 & 2071 & 1907 & 1482.220 & 1484.896 & 0 & \\
43-2-1 & 4172 & 4098 & 1487.181 & 1492.960 & 3 & \\
43-2-2 & 3855 & 3481 & 1492.971 & 1497.875 & 1 & Removed 0.25 day from end\\
44-1-1 & 2832 & 2714 & 1500.281 & 1504.107 & 2 & \\
44-1-2 & 2790 & 2751 & 1504.118 & 1507.982 & 0 & \\
44-1-3 & 2689 & 2663 & 1507.993 & 1511.727 & 1 & \\
44-2-1 & 2356 & 1820 & 1514.420 & 1516.983 & 1 & Removed 0.50 day from start\\
44-2-2 & 2698 & 2666 & 1516.994 & 1520.733 & 1 & \\
44-2-3 & 2664 & 2619 & 1520.744 & 1524.443 & 0 & \\
45-1-1 & 4264 & 3705 & 1526.584 & 1531.796 & 2 & Removed 0.50 day from start \\
45-1-2 & 4447 & 4390 & 1531.808 & 1537.983 & 0 & \\
45-2-1 & 3486 & 3286 & 1540.165 & 1544.797 & 4 & \\
45-2-2 & 4192 & 4126 & 1544.808 & 1550.629 & 1 & \\
\enddata
\textbf{Notes.}
\vspace{-0.2cm}\tablenotetext{\textrm{a}}{The numbers indicate the TESS sector, spacecraft orbit (two per sector), and segment number, respectively.}
\vspace{-0.2cm}\tablenotetext{\textrm{b}}{Number of datapoints in each data segment before and after removing flagged points, outliers, and flux ramps.}
\vspace{-0.2cm}\tablenotetext{\textrm{c}}{Start and end times of each data segment ($\mathrm{BJD}_{\mathrm{TDB}}-2{,}458{,}000$).}
\vspace{-0.2cm}\tablenotetext{\textrm{d}}{Order of the polynomial detrending model used in the full-orbit phase-curve fitting.}
\vspace{-0.9cm}
\end{deluxetable*}

\begin{figure*}[t!]
\centering
\includegraphics[width=0.8\linewidth]{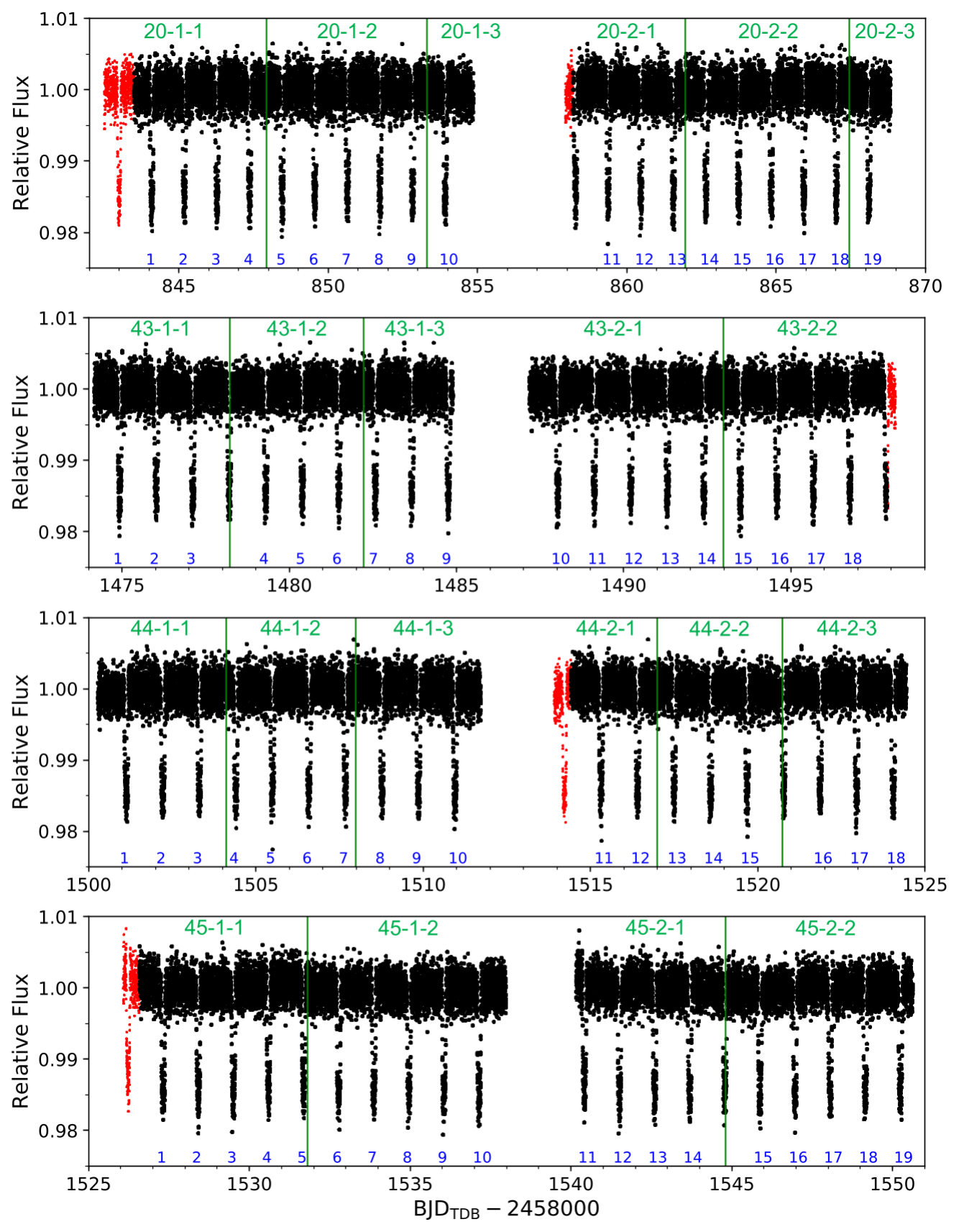}
\caption{Normalized and outlier-removed TESS PDC light curves of WASP-12 from Sectors 20, 43, 44, and 45. The red dots indicate portions of the photometry that were trimmed prior to fitting due to short-term flux ramps. The vertical green lines denote the scheduled momentum dumps, and each data segment is labeled with the corresponding sector, orbit, and segment numbers. The primary transits that were included in the transit-timing analysis are labeled sequentially in blue.}
\label{fig:raw}
\end{figure*}

TESS first observed WASP-12 (TIC 86396382) during Sector 20 of the primary mission, which lasted from UT 2019 December 24 to 2020 January 21. The  WASP-12 data were obtained with Camera 1 in the two-minute cadence mode. The data were downlinked from the spacecraft and passed through the Science Processing Operations Center (SPOC) pipeline hosted at the NASA Ames Research Center \citep{jenkins2016}. The pipeline determined the optimal photometric aperture and extracted the light curve, which was deposited in the Mikulski Archive for Space Telescopes (MAST).

Analyses of the Sector 20 light curves of WASP-12 have been published by several teams. \citet{turner2021} analyzed the transits and secondary eclipses and used the measured timings to calculate a refined orbital decay rate for WASP-12b. The full-orbit phase curve of the WASP-12 system was analyzed independently by \citet{owens2021} in a standalone paper and by \citet{wong2021year2} as part of a systematic phase-curve study of short-period exoplanet systems observed during the second year of the TESS mission.

TESS revisited the northern ecliptic hemisphere beginning in mid-2021. Due to the fortuitous positioning of the fields of view, WASP-12 was observed in three consecutive sectors: Sector 43 (2021 September 16 to October 12), Sector 44 (2021 October 12 to November 6), and Sector 45 (2021 November 6 to December 2). During these three roughly month-long observations, the target was located on Cameras 4, 3, and 1, respectively. The combined TESS dataset for WASP-12 now spans almost two years.

In the analysis of the TESS WASP-12 photometry presented in this paper, we utilized the pre-search data conditioning (PDC) light curves, which were corrected for common-mode instrumental systematic trends by the SPOC pipeline using empirically derived cotrending basis vectors \citep{smith2012,stumpe2012,stumpe2014}. Our experience with TESS light-curve fitting has demonstrated that the PDC light curves typically exhibit lower levels of time-correlated noise when compared to the raw simple aperture photometry (SAP) light curves, while preserving genuine astrophysical variability, including the orbital phase curve (see, for example, \citealt{wong2020wasp19,wong2020kelt9}).

The preprocessing steps that we completed prior to fitting were largely identical to the methodology described in our previous TESS light-curve papers \citep[e.g.,][]{wong2020year1,wong2021year2,wong2021toi2109}. First, we removed all flagged points and trimmed $3\sigma$ outliers from the transit-masked photometry using a 16-point-wide moving median filter. Next, we separated the light curves into data segments that are separated by the scheduled momentum dumps and data downlink interruptions. Lastly, we searched for significant flux ramps that often occur at the beginning or end of data segments and removed them from the time series. The list of TESS light-curve segments is provided in Table~\ref{tab:segments}, including the number of datapoints, time ranges, and any trimmed portions. Figure~\ref{fig:raw} shows the PDC light curves from the four sectors of the TESS observations, with the momentum dumps and trimmed ramps marked.

\subsection{Palomar/WIRC Secondary Eclipse}\label{subsec:wirc}

We observed a full secondary eclipse of WASP-12b on UT 2021 February 18 with the Wide-field Infrared Camera (WIRC; \citealt{wilson2003}) on the Hale $200''$ Telescope at Palomar Observatory, California, USA. Images were taken in the $K_s$ band ($\lambda_{\rm eff} = 2.13$ $\mu$m) using a special beam-shaping diffuser for improved duty cycle and guiding stability; see \citet{stefansson2017} and \citet{vissapragada2020} for details concerning the observing setup. Following an initial two-point dither near the target to construct a background frame and mitigate short-cadence detector systematics, we obtained 178 exposures with a total integration time of 42 s (14 coadds $\times$ 3 s).

After dark-correcting, flat-fielding, and background-subtracting each image using the methods described by \citet{vissapragada2020}, we extracted the photometry of WASP-12 and 10 nearby companion stars using circular apertures ranging in diameter from 10 to 25 pixels (in 1 pixel steps). For our subsequent analysis, we chose a 17 pixel aperture, which was found to minimize the per-point scatter in the resultant WASP-12 light curve.

\section{Data Analysis}\label{sec:ana}

For our analysis of the TESS photometry, we used the Python-based \texttt{ExoTEP} pipeline \citep[e.g.,][]{benneke2019,wong2020hatp12}. In this section, we provide a brief overview of our fitting methodology for the transits, secondary eclipses, and full-orbit phase curves. We refer the reader to previous studies, such as \citet{shporer2019} and \citet{wong2020year1}, for more detailed descriptions of the light-curve modeling and error analysis.

\subsection{Transits}\label{subsec:transits}

We considered all 74 full-transit events in the trimmed TESS light curves that were not interrupted by a momentum dump: 19 in Sector 20, 18 in Sector 43, 18 in Sector 44, and 19 in Sector 45. These transits are labeled sequentially within each sector in Figure~\ref{fig:raw}. To construct each individual transit time series, we selected all datapoints with timestamps within $0.1P$ of the predicted mid-transit time, where $P$ is the orbital period. There were about 160 selected points on average. This procedure ensured a sufficient out-of-transit baseline, while simultaneously minimizing the effect of the orbital phase-curve variation on the fitted transit depth.

Within the \texttt{ExoTEP} pipeline, transit light-curve modeling is handled by \texttt{batman} \citep{batman}, which is based on the analytical formulation of \citet{mandel2002}. For every transit, the unconstrained free parameters were the mid-transit time $T_c$, planet--star radius ratio $R_p/R_*$, impact parameter $b$, and scaled semimajor axis $a/R_*$. The orbital period $P$ was held fixed at the extrapolated value based on the quadratic transit ephemeris reported by \citet{turner2021}. We assumed a quadratic limb-darkening law with coefficients constrained by Gaussian priors: the median values ($u_1 = 0.24$, $u_2 = 0.31$) were calculated by interpolating the coefficients tabulated by \citet{claret2017} based on the measured properties of the host star WASP-12 (from Version 8 of the TESS Input Catalog; \citealt{stassun2019}), and the Gaussian widths were set to 0.05 to account for the uncertainties in the stellar parameters.

Parameter estimation was carried out using the affine-invariant Markov Chain Monte Carlo (MCMC) ensemble sampler \texttt{emcee} \citep{emcee}. In addition to the free parameters listed above, we included a per-point uncertainty parameter $\sigma$ that was allowed to vary freely to ensure that the resultant reduced $\chi^2$ was at unity. For normalization and systematics detrending, we multiplied the transit model by either a constant multiplicative constant or a linear function of time. We utilized the Bayesian information criterion (BIC) to determine the optimal systematics model on a transit-by-transit basis; the linear trend was preferred in only one case --- the second transit of Sector 43 (S43-2).

The results of our individual transit fits are listed in Table~\ref{tab:transits} in the Appendix. The measured transit depths show a very high level of mutual consistency, with the largest pairwise discrepancy being $2.3\sigma$. Likewise, the transit-shape parameters $b$ and $a/R_*$ show little variation across the 74 transits, with the spread in values lying within $2.4$ and $2.5\sigma$, respectively. The median $1\sigma$ timing uncertainty for the TESS transits is 42 s. Compilation plots of the normalized transit light curves and best-fit models are provided in Figures~\ref{fig:trans20}--\ref{fig:trans45}.

\subsection{Secondary Eclipses}\label{subsec:eclipses}

The individual secondary eclipse events are not detected with a sufficiently high signal-to-noise ratio to permit reliable timing measurements. Instead, we combined all full and uninterrupted occultation events within a TESS sector and calculated a single effective mid-occultation time. In an analogous manner to our transit analysis, we selected points within 10\% of an orbit from the predicted mid-occultation time. After median-normalizing each secondary eclipse event, we phase-folded the datapoints. For each sector, we held $P$, $R_p/R_*$, $b$, and $a/R_*$ fixed to the best-fit values determined from the corresponding individual sector full-orbit phase-curve fit (Section~\ref{subsec:phase}, Table~\ref{tab:sectors}). 

We fit for the mid-occultation time $T_{\rm occ}$ and secondary eclipse depth $D_{d}$, along with the uniform per-point uncertainty $\sigma$. The epoch of $T_{\rm occ}$ for each sector was set to the secondary eclipse event closest to the midpoint of the time series. The best-fit occultation timings for the four TESS sectors are (in ${\rm BJD}_{\rm TDB}$)
\begin{align}
    T_{\rm occ,S20} = 2{,}458{,}855.5543 \pm 0.0029, \notag \\
    T_{\rm occ,S43} = 2{,}459{,}485.3066 \pm 0.0021, \notag \\
    T_{\rm occ,S44} = 2{,}459{,}511.4976 \pm 0.0026, \notag \\
    T_{\rm occ,S45} = 2{,}459{,}537.6908 \pm 0.0020. \label{eq:ecl1}
\end{align}
The corresponding measured secondary eclipse depths are $534 \pm 54$, $497 \pm 49$, $450 \pm 51$, and $498 \pm 46$ ppm, respectively. Later in this paper, we will present improved measurements of the occultation depths derived from the full-orbit phase-curve fits. The improvement comes from a more robust treatment of the time-correlated noise and self-consistently accounting for the out-of-eclipse phase-curve modulations as well as the uncertainties in the orbital parameters.

\begin{figure}[t!]
\centering
\includegraphics[width=\linewidth]{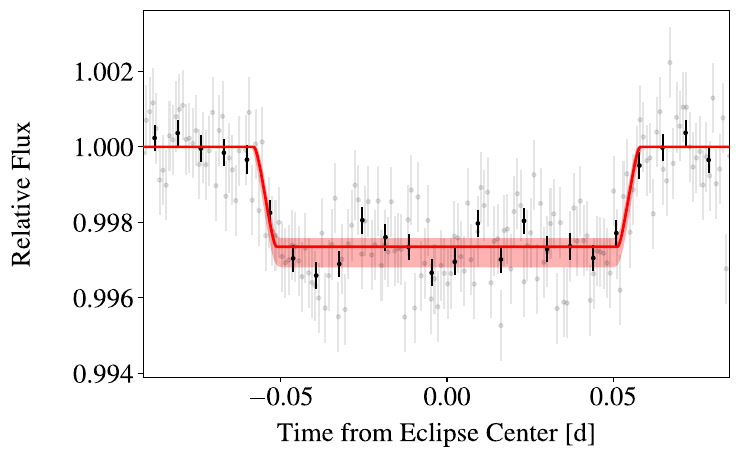}
\caption{The normalized and detrended $K_s$-band secondary eclipse light curve observed with Palomar/WIRC on UT 2021 February 18 (gray). The black datapoints show the light curve binned in 10 minute intervals. The red curve and shaded area denote the best-fit occultation model and $1\sigma$ confidence region.}
\label{fig:kband}
\vspace{-0.4cm}
\end{figure}

We fit the Palomar/WIRC $K_s$-band secondary eclipse light curve with the \texttt{exoplanet} toolkit \citep{exoplanet}. For detrending vectors, we used the extracted flux arrays from the 10 companion stars, airmass, background flux, and the relative centroid offset of WASP-12. These vectors were combined in linear combination with a constant multiplicative offset, and the coefficients were included in the MCMC fit as free parameters along with the mid-occultation time and eclipse depth. We applied Gaussian priors to the remaining orbital parameters based on the measured values from the previously published Sector 20 phase-curve fit in \citet{wong2021year2}.

We obtained a mid-occultation time of 
\begin{equation}
T_{{\rm occ, WIRC}} = 2{,}459{,}263.74611 \pm 0.00090 \,\,{\rm BJD}_{\rm TDB}
\label{eq:ecl2}
\end{equation}
and an eclipse depth of $2810 \pm 390$ ppm. The precision of this timing measurement is significantly higher than that of the TESS occultation timings. The normalized and detrended $K_s$-band secondary eclipse light curve is plotted in Figure~\ref{fig:kband}, along with the best-fit occultation model and $1\sigma$ confidence region.

\subsection{Full-orbit Phase-curve Fits}\label{subsec:phase}

% Individual sector results
%-------------------------------------------------------------------------------
\begin{deluxetable*}{lllll|l}
\tablewidth{0pc}
\setlength{\tabcolsep}{3pt}
\renewcommand{\arraystretch}{0.95}
\tabletypesize{\footnotesize}
\tablecaption{
    Results from Individual Sector Phase-curve Fits and Combined Analysis
    \label{tab:sectors}
}
\tablehead{  & & & & & \\[-0.3cm] \multicolumn{1}{l}{Parameter} &
   \multicolumn{1}{c}{Sector 20} & \multicolumn{1}{c}{Sector 43} &
\multicolumn{1}{c}{Sector 44} & \multicolumn{1}{c}{Sector 45} & \multicolumn{1}{c}{All}
}
\startdata
\multicolumn{2}{l}{Fitted Parameters} & & & \\
$R_p/R_*$     & $0.11666 \pm 0.00081$ & $0.1152 \pm 0.0010$ & $0.11581 \pm 0.00079$ & $0.11563 \pm 0.00070$ & $0.11600 \pm 0.00045$\\
$T_c$\tablenotemark{\scriptsize a}       & $855.01067 \pm 0.00013$ & $1484.75805 \pm 0.00010$ & $1510.95210 \pm 0.00012$ & $1537.14639 \pm 0.00014$ & $1196.624116 \pm 0.000072$\\
$P$ (days)    & $1.091408 \pm 0.000019$ & $1.091433 \pm 0.000015$ & $1.091407 \pm 0.000016$ & $1.091401 \pm 0.000019$ & $1.09141685 \pm 0.00000025$ \\
$b$  & $0.311_{-0.094}^{+0.069}$ & $0.327_{-0.091}^{+0.067}$ & $0.357_{-0.077}^{+0.050}$ & $0.293_{-0.093}^{+0.071}$ & $0.344 \pm 0.036$ \\
$a/R_*$       & $3.086 \pm 0.070$ & $3.080_{-0.070}^{+0.071}$ & $3.048_{-0.056}^{+0.068}$ & $3.107_{-0.073}^{+0.068}$ & $3.061 \pm 0.034$ \\
$\bar{f_p}$ (ppm) & $197 \pm 74$ & $217 \pm 61$ & $197 \pm 75$ & $201 \pm 73$ & $205 \pm 31$ \\
$A_{\mathrm{atm}}$ (ppm)   & $265 \pm 36$ & $305 \pm 27$ & $223 \pm 26$ & $249 \pm 37$ & $262 \pm 15$ \\
$\delta$ (deg) & $10.3 \pm 6.3$ & $5.8 \pm 4.0$ & $3.5 \pm 5.3$ & $6.3 \pm 6.3$ & $6.2 \pm 2.8$ \\
$A_{\mathrm{ellip}}$ (ppm)  & $77 \pm 38$ & $102 \pm 30$ & $52 \pm 32$ & $30 \pm 38$ & $65 \pm 14$ \\
$u_1$\tablenotemark{\scriptsize b} & $0.22 \pm 0.03$ & $0.23 \pm 0.03$ & $0.22 \pm 0.03$ & $0.23 \pm 0.03$ & $0.22 \pm 0.02$ \\
$u_2$\tablenotemark{\scriptsize b} & $0.30 \pm 0.05$  & $0.32 \pm 0.05$ & $0.30 \pm 0.05$ & $0.28 \pm 0.05$ & $0.30 \pm 0.04$ \\
\multicolumn{2}{l}{Derived Parameters} & & & \\
$D_{d}$ (ppm)\tablenotemark{\scriptsize c}  & $454 \pm 80$ &  $519 \pm 67$ & $418 \pm 83$ & $450 \pm 79$ & $466 \pm 35$ \\
$D_{n}$ (ppm)\tablenotemark{\scriptsize c}   & $-64 \pm 83$ & $-89 \pm 69$ & $-25 \pm 79$ & $-46 \pm 81$ & $-55 \pm 35$ \\
$i$ (deg)      & $84.2^{+1.8}_{-1.4}$ & $83.9^{+1.8}_{-1.4}$ & $83.3^{+1.6}_{-1.1}$ & $84.5^{+1.8}_{-1.5}$ & $83.54 \pm 0.74$ \\
\enddata
\textbf{Notes.}
\vspace{-0.2cm}\tablenotetext{\textrm{a}}{${\rm BJD}_{\rm TDB} - 2{,}458{,}000$.}
\vspace{-0.2cm}\tablenotetext{\textrm{b}}{The quadratic limb-darkening coefficients were constrained by Gaussian priors based on the tabulated values in \citet{claret2017}: $u_1 = 0.24 \pm 0.05$, $u_2 = 0.31 \pm 0.05$.}
\vspace{-0.2cm}\tablenotetext{\textrm{c}}{$D_{d}$ and $D_{n}$ denote the dayside and nightside fluxes, respectively. The dayside flux is equivalent to the secondary eclipse depth.}
\vspace{-0.9cm}
\end{deluxetable*}

The phase curve consists of the transits, secondary eclipses, and the synchronous brightness modulation of the entire system. The flux variability contains contributions from the changing viewing geometry of the planet's longitudinal atmospheric brightness modulation as it orbits around the host star, as well as the ellipsoidal distortion of both orbiting companions due to the mutual gravitational interaction and the Doppler boosting of the star's spectrum (see the review by \citealt{shporer2017}). In tandem with previous TESS phase-curve studies, we used the following simple sinusoidal light-curve model \citep[e.g.,][]{shporer2019,wong2020wasp19,wong2020year1,wong2020kelt9,wong2021year2,wong2021toi2109}:
\begin{align}
\label{astro}F(t) &= \frac{F_{*}(t)\lambda_t(t) + F_{p}(t)\lambda_e(t)}{1 + \bar{f_p}},\\
\label{planet}F_{p}(t) &= \bar{f_{p}} - A_{\rm atm} \cos(\phi + \delta),\\
\label{star}F_{*}(t) &= 1 - A_{\rm ellip}\cos(2\phi) + A_{\rm Dopp}\sin(\phi),\\
\phi &\equiv 2\pi(t - T_c).
\end{align}
Here, $\lambda_t$ and $\lambda_e$ denote the unit-normalized transit and secondary eclipse light curves. The flux from the planet, $F_{p}$, is parameterized by the average flux level $\bar{f_{p}}$, the semi-amplitude of the atmospheric brightness modulation $A_{\rm atm}$, and the phase shift in the planet's phase curve $\delta$. The secondary eclipse depth (i.e., dayside hemispheric brightness) is derived from the aforementioned three parameters via $D_d = \bar{f_{p}} - A_{\rm atm}\cos(\pi + \delta)$. Similarly, the nightside flux is given by $D_n = \bar{f_{p}} - A_{\rm atm}\cos{\delta}$. The stellar flux $F_{*}$ includes two harmonic amplitudes, $A_{\rm ellip}$ and $A_{\rm Dopp}$, which correspond to the ellipsoidal distortion and Doppler boosting signals \citep[e.g.,][]{loeb2003,faigler2011,faigler2015,shporer2017}.

The PDC light curves are affected by low-level residual systematics in the form of long-term temporal trends. To model these systematics, we multiplied the astrophysical phase-curve model by polynomial functions of time, with a separate systematics function defined for each data segment. We carried out preliminary phase-curve fits to each data segment and determined the optimal polynomial order using the BIC. These selected orders are listed in Table~\ref{tab:segments} and are predominantly 0 or 1.

Each sector's light curve was analyzed separately using a two-step fitting process. All of the orbital parameters ($T_c$, $P$, $b$, and $a/R_*$) and most transit/eclipse/phase-curve parameters ($R_p/R_*$, $\bar{f_{p}}$, $A_{\rm atm}$, $\delta$, and $A_{\rm ellip}$) were allowed to vary freely. We accounted for the light-travel time across the system when modeling the secondary eclipses by adding $2a/c \approx 23$ s to the calculated time of superior conjunction. Our model neglects the gradual decrease in the orbital period over the timescale of a TESS sector, because the expected transit-timing deviation is only about 6 ms. The limb-darkening coefficients were constrained by the same Gaussian priors as in our transit fits ($u_1 = 0.24 \pm 0.05$, $u_2 = 0.31 \pm 0.05$). Meanwhile, given the small amplitude of the Doppler boosting signal and its degeneracy with the planet's phase-curve offset $\delta$, we followed the technique used by \citet{wong2020year1} and \citet{wong2021year2} and applied a Gaussian prior to $A_{\rm Dopp}$. The median and width of this prior was derived from the theoretical formulation described by \citet{shporer2017} and the system parameters published by \citet{stassun2019}: $2.3 \pm 0.2$ ppm.

In the first MCMC fit, we allowed the per-point uncertainty $\sigma$ to vary. The best-fit $\sigma$ values for the four TESS sectors were 1880, 1790, 1820, and 1780 ppm, respectively. Upon examination of the residuals, we found discernible time-correlated noise (i.e., red noise). To propagate this red-noise contribution to the best-fit phase-curve parameters and uncertainties, we computed the scatter of the residuals binned at intervals ranging from 20 minutes to 8 hr and calculated the average fractional deviation $\beta$ from the expected $1/\sqrt{n}$ trend for pure white noise (see \citealt{wong2020year1} for a full description of this methodology). We then multiplied the best-fit $\sigma$ from the first fit by $\beta$ to obtain the red-noise-enhanced per-point uncertainty $\sigma_r\equiv\beta\sigma$. For the second phase-curve fit, we fixed the flux uncertainties to $\sigma_r$ and reran the MCMC analysis to arrive at the final set of results. Across the four TESS sectors, $\beta$ ranged from 1.01 to 1.23, illustrating the small and variable contribution of red noise throughout the light curves.

The results of our individual sector phase-curve fits are listed in Table~\ref{tab:sectors}. There is a high level of consistency in the measured astrophysical parameters across the four sectors. The planet--star radius ratios agree at better than the $1.1\sigma$ level, and the transit-shape parameters ($b$, $a/R_*$) are mutually consistent to well within $1\sigma$. Meanwhile, the secondary eclipse depths $D_d$ and atmospheric brightness modulation amplitudes $A_{\rm atm}$ lie within $1.0$ and $2.2\sigma$, respectively. Likewise, the planet's phase-curve offsets $\delta$ are in excellent agreement, and the nightside fluxes are all consistent with zero. The star's ellipsoidal distortion signal amplitudes show a $1.3\sigma$ spread. All in all, there is no sign of statistically significant alterations in the shape of the WASP-12 phase curve throughout the TESS observations. Constraints on atmospheric variability are discussed in more detail in Section~\ref{subsec:variability}.

Having established that the transits, secondary eclipses, and synchronous photometric modulations of the WASP-12 system are comparable across the four TESS sectors, we proceeded with a joint phase-curve fit to the combined TESS light curve. Just as in our individual sector phase-curve analyses, we assumed the orbital period to be constant throughout the time interval of the TESS observations. Choosing the reference epoch to be the transit event nearest to the midpoint of the full TESS time series, the accumulated transit-time deviation due to the shrinking period is expected to be less than 15 s at the first and last epochs. A deviation of 15 s is shorter than both the 120 s cadence of the TESS observations and the average precision with which the individual mid-transit times are measured (42 s). Prior to fitting, we divided away the best-fit systematics models from the light curve (calculated in the individual sector phase-curve fits) and fixed the per-point uncertainties to the respective red-noise-enhanced $\sigma_r$ values that we computed previously.

\begin{figure}[t]
\centering
\includegraphics[width=\linewidth]{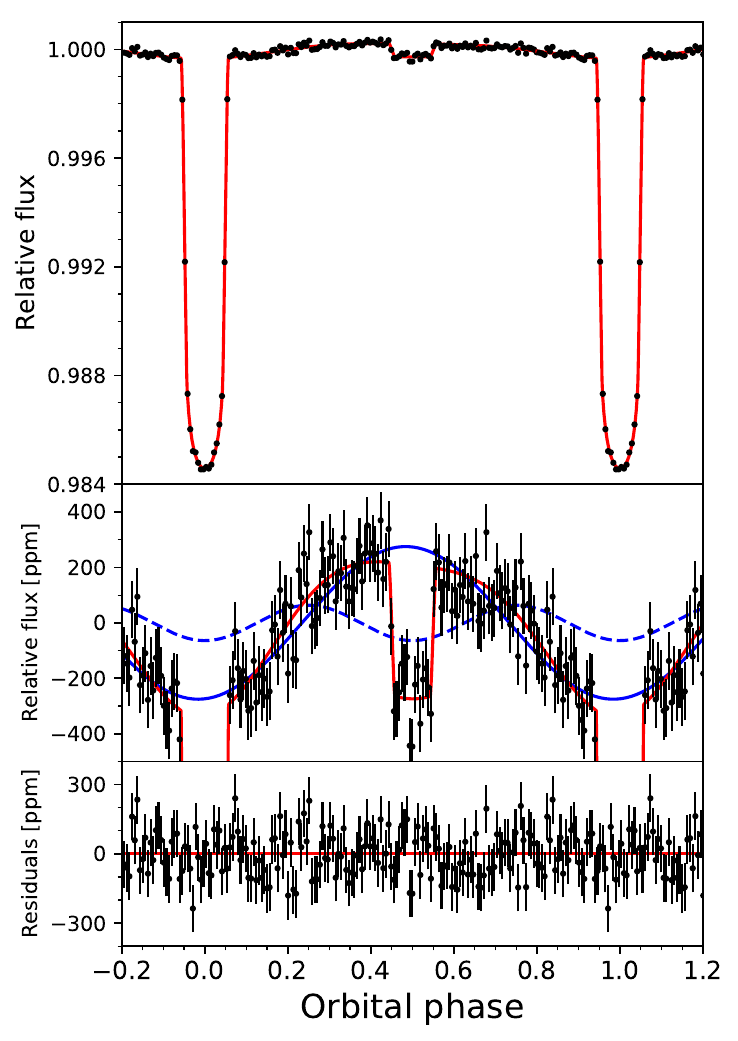}
\caption{Top: the combined four-sector TESS phase curve, phase-folded and binned in 10-minute intervals. The red curve shows the best-fit full-orbit phase-curve model. Middle: a zoomed-in view of the phase-curve variability. The solid and dashed blue curves denote the individual components of the phase curve that correspond to atmospheric brightness modulation and ellipsoidal distortion, respectively. Bottom: the residuals from the best-fit model.}
\label{fig:fullphase}
\vspace{-0.4cm}
\end{figure}

The results of our combined phase-curve analysis are listed in the rightmost column of Table~\ref{tab:sectors}. The phase-folded light curve and best-fit model are shown in Figure~\ref{fig:fullphase}. The fitted values yield a more than twofold improvement in precision when compared to the previously published Sector 20 phase-curve fit results in \citet{wong2021year2}. The transit-shape parameters were tightly constrained: $b = 0.344 \pm 0.036$, $a/R_* = 3.061 \pm 0.034$, and $i = 83\overset{\circ}{.}54 \pm 0\overset{\circ}{.}74$. The planet--star radius ratio of $R_p/R_* = 0.11600 \pm 0.00045$ is among the most precise in the literature. Using the stellar radius $R_* = 1.75 \pm 0.09\,R_{\Sun}$ published in \citet{stassun2019}, we find $R_p = 1.98 \pm 0.10\,R_{\rm Jup}$ and $a = 0.0249 \pm 0.0013$ au. 

The TESS-band secondary eclipse depth of WASP-12b is $466 \pm 35$ ppm, and the nightside flux is consistent with zero at $1.6\sigma$ ($2\sigma$ upper limit of 70 ppm). The day--night brightness contrast has a peak-to-peak amplitude of $524 \pm 30$ ppm. There is a small, marginally significant $6\overset{\circ}{.}2 \pm 2\overset{\circ}{.}8$ eastward shift in the location of maximum dayside brightness. Given the low optical geometric albedo of WASP-12b \citep{bell2017,wong2021year2}, the planetary phase curve observed in the TESS bandpass is dominated by the thermal emission from the atmosphere. It follows that the measured phase offset suggests the presence of superrotating equatorial winds that advect heat toward the colder nightside. 

The star's ellipsoidal distortion signal has a semi-amplitude of $65 \pm 14$ ppm. Using the theoretical relationship between the ellipsoidal distortion amplitude and the planet--star mass ratio \citep[e.g.,][]{shporer2017}, along with the measured system parameters and coefficients for limb-darkening and gravity-darkening interpolated from the tables in \citet{claret2017}, we arrived at an independent estimate of WASP-12b's mass: $1.96 \pm 0.57\,M_{\rm Jup}$. This value is consistent with the planet mass presented in the discovery paper ($1.41 \pm 0.10\,M_{\rm Jup}$; \citealt{wasp12}) to within $1\sigma$. The close agreement indicates that the photometric modulation at the first harmonic of the orbital frequency is adequately modeled by the classical formalism of stellar tidal distortion.

A recent radial velocity (RV) analysis of the WASP-12 system demonstrated that the apparent signature of a small orbital eccentricity could be explained instead by the flow velocity of the stellar atmosphere due to the tidal bulge raised by the orbiting planet \citep{m20b}. The measured tidal RV amplitude was $7.3\,{\rm m}\,{\rm s}^{-1}$, which is expected to produce a photometric modulation with a semi-amplitude of roughly 80 ppm, in good agreement with our measured ellipsoidal distortion signal.

\section{Discussion}\label{sec:dis}

\subsection{Orbital Decay}\label{subsec:decay}

\begin{figure*}[t!]
\centering
\includegraphics[width=\linewidth]{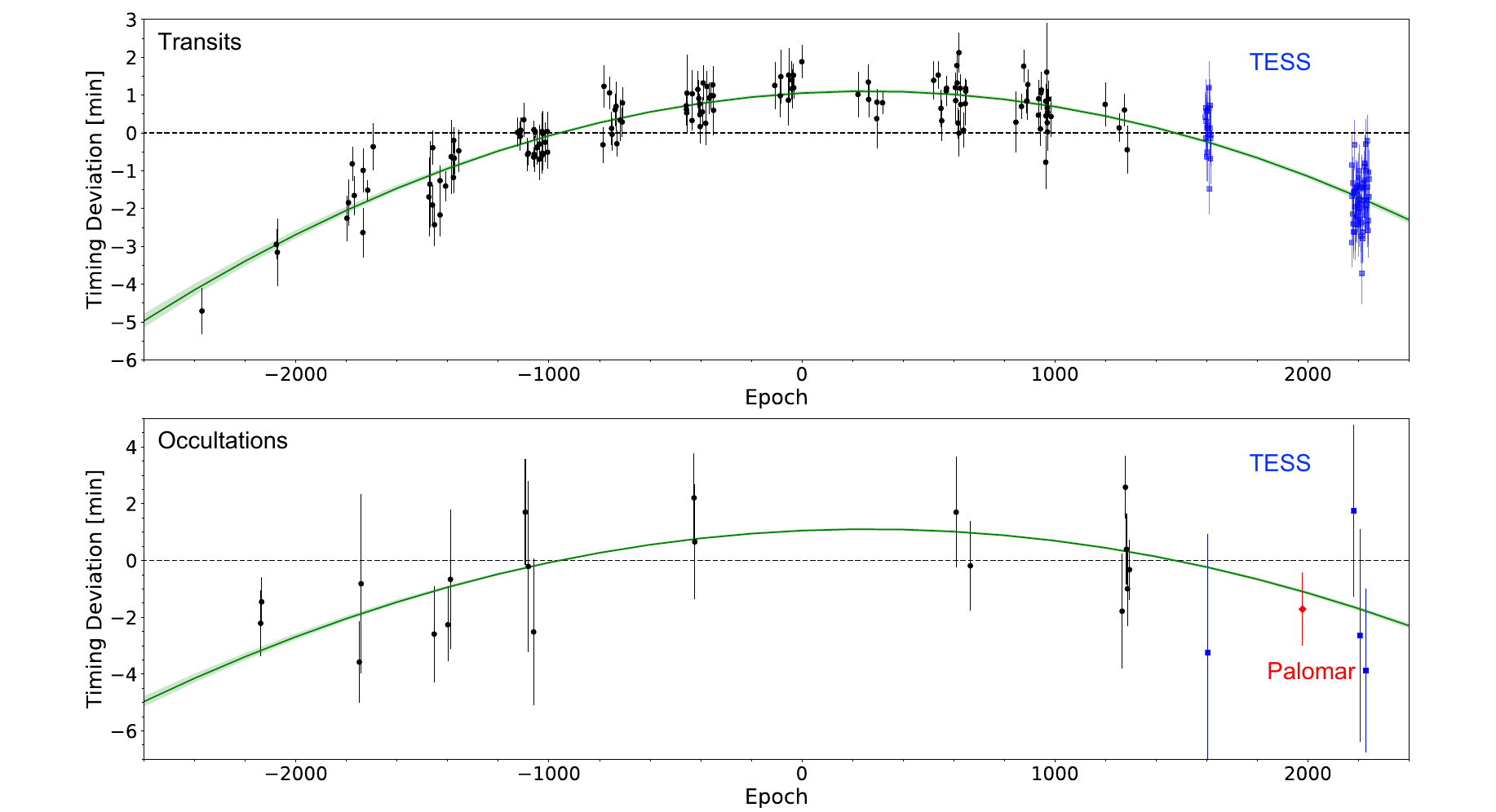}
\caption{Plot of the transit (top) and occultation (bottom) timing deviations from the best-fit constant-period orbital ephemeris. The black points are previously published data from \citet{yee2020}, while the blue points denote the newly measured timings from our TESS light-curve analysis (Table~\ref{tab:transits} and Section~\ref{subsec:eclipses}). The red datapoint is based on the $K_s$-band Palomar/WIRC secondary eclipse observation. The green curves and shaded areas indicate the best-fit orbital decay ephemeris model and corresponding $1\sigma$ confidence regions.}
\label{fig:ephem}
\end{figure*}

Previous transit-timing analyses \citep{patra2017,m18a,yee2020,turner2021} uncovered strong evidence for a shrinking orbital period. Having obtained a large number of new transit timings from TESS Sectors 43--45 and extended the overall time baseline by almost 2 years, we carried out an updated ephemeris fit to refine the orbital decay-rate estimate.

We combined the TESS transit timings listed in Table~\ref{tab:transits} and the five mid-occultation timings presented in Equations~\eqref{eq:ecl1} and \eqref{eq:ecl2} with the full list of prior transit and occultation times compiled by \citet{yee2020}.\footnote{The previous transit and secondary eclipse timings were contributed by \citet{wasp12}, \citet{campo2011}, \citet{chan2011}, \citet{cowan2012}, \citet{crossfield2012}, \citet{sada2012}, \citet{copperwheat2013}, \citet{fohring2013}, \citet{m13a}, \citet{stevenson2014b}, \citet{croll2015}, \citet{deming2015}, \citet{kreidberg2015}, \citet{m16}, \citet{collins2017}, \citet{patra2017},  \citet{m18a}, \citet{hooton2019}, \citet{ozturk2019}, \citet{vonessen2019}, \citet{yee2020}, and \citet{patra2020}.} There are a total of 213 transit timings and 24 occultation timings. The transit- and occultation-time models for orbital decay are
\begin{align}
T_{\rm tra}(E) &= T_{0}+PE+\frac{1}{2}\frac{dP}{dE}E^2, \notag\\
T_{\rm occ}(E) &= T_{0}+P\left(E+\frac{1}{2}\right)+\frac{1}{2}\frac{dP}{dE}E^2. \label{quad}
\end{align}
The variable $dP/dE$ represents the decay rate in units of days per orbit, $T_0$ is the zeroth epoch transit time, and $E$ is the transit epoch.

Before fitting the measured timings with \texttt{emcee}, we subtracted the light-travel time of 23 s from the mid-occultation times. We designated the transit event closest to the weighted average of timings as the zeroth epoch in our fit. The results of our MCMC analysis are
\begin{align}
T_{0} &= 2{,}457{,}103.283654 \pm 0.000032 \,\, {\rm BJD}_{\rm TDB}, \notag\\
P &= 1.091419370 \pm 0.000000020 \,\, {\rm days},\notag \\
\frac{dP}{dE} &= (-1.031 \pm 0.033) \times 10^{-9} \, {\rm days}~ {\rm orbit}^{-1}, \notag\\
\frac{dP}{dt} &= -29.81 \pm 0.94 \,\, {\rm ms}~{\rm yr}^{-1}. \label{eqn:ephem}
\end{align}
The updated orbital decay timescale is $\tau = P/|\dot{P}| = 3.16 \pm 0.10\,{\rm Myr}$, which lies in between the published estimates of \citet{yee2020} and \citet{turner2021}. Figure~\ref{fig:ephem} shows the timing deviations of the WASP-12b transits and secondary eclipses relative to a constant-period ephemeris model, with the best-fit orbital decay model and corresponding $1\sigma$ confidence regions plotted in blue. The fit quality is excellent, with a reduced $\chi^2$ of 1.05 and 234 degrees of freedom (237 datapoints and 3 free parameters).

The additional time observations from the TESS extended mission improved the precision of the measured orbital decay rate by almost 30\% when compared to the estimate by \citet{turner2021}, which incorporated timings from Sector 20 only. The magnitude of our derived decay rate ($29.81 \pm 0.94$ ms~yr$^{-1}$) is $1.5\sigma$ larger than the previously published value ($27.3 \pm 1.4$ ms~yr$^{-1}$).\footnote{We note that the transformation from $dP/dE$ to $dP/dt$ in \citet{turner2021} is erroneous: the orbital period conversion factor 1.09 days orbit$^{-1}$ was multiplied, instead of divided.} To assess the influence of the occultation timings on the measured orbital decay rate, we carried out an analogous fit without the occultation timings and obtained $30.01 \pm 0.96$ ms~yr$^{-1}$. This indicates that while the occultations are crucial for distinguishing between the orbital decay and apsidal precession scenarios \citep{yee2020,turner2021}, the timings do not have a substantial impact on the decay-rate estimate.

The measured orbital decay rate $dP/dt$ can be used to constrain the host star's modified\footnote{$Q'_*$ is defined as $3Q_*/2k_2$, where $Q_*$ is the tidal quality factor and $k_2$ is the second-order Love number.} tidal quality factor $Q'_*$, which in the model of \citet{goldreich1966} is related to the period derivative via
\begin{equation}
\label{eq:dpdt}
    \left|\frac{dP}{dt}\right| = \frac{27\pi q}{2Q'_*}\left(\frac{R_*}{a}\right)^5.
\end{equation}
Using the planet--star mass ratio $q\equiv M_p/M_*$ from \citet{collins2017} and $a/R_*$ from our combined phase-curve fit (Table~\ref{tab:sectors}), we arrived at $Q'_* = (1.50 \pm 0.11)\times 10^5$.

As has been discussed in the previous orbital decay studies \citep{yee2020,turner2021}, the measured $Q'_*$ value of WASP-12 lies at the lower end of most other estimates in the literature based on theoretical calculations or less direct empirical methods. The range of tidal quality factors inferred from analyzing the eccentricity distribution of main-sequence stellar binaries is $10^5$--$10^7$ \citep[e.g.,][]{ogilvie2007,lanza2011,meibom2015}, while analogous population studies of hot-Jupiter hosts point toward $Q'_*$ values in the range of $10^{5.5}$--$10^{6.5}$ \citep[e.g.,][]{jackson2008,husnoo2012,barker2020}. Other population-level approaches indicate a broad range of possible tidal quality factors ($10^5$--$10^8$; \citealt{cameron2018,penev2018,hamer2019}). 

Meanwhile, the theoretically expected values of $Q'_*$ are generally higher ($10^7$--$10^{10}$; see \citealt{ogilvie2014} and references therein). \citet{weinberg2017} proposed that WASP-12 is a subgiant star for which more rapid tidal dissipation might be expected, due to breaking of internal gravity waves near the star's core. This hypothesis was investigated further by \citet{bailey2019}, who agreed with the theoretical premise but found that the observed characteristics of WASP-12 made it more likely to be a main-sequence star. \citet{barker2020} also calculated tidal dissipation rates for WASP-12 and demonstrated that the theoretical and observed decay rates could only be brought into agreement if the star is a subgiant. Taking a different track, \citet{millholland2018} proposed that the orbital decay is driven by tidal dissipation within the planet, rather than the star; this would be due to a hypothetical nonzero planetary obliquity that is maintained by a nearby planetary companion. This intriguing scenario was investigated further by \citet{su2021}, who concluded that it required too much fine tuning to be plausible.

In short, there is not yet a completely satisfactory explanation for the rapid orbital decay of WASP-12b. Some priorities for future progress in this realm are better characterization of the host star and its evolutionary state, as well as more sensitive searches for nearby planetary companions.
%We mention in passing that the $Q'_*$ value of XO-3, calculated based on the assumption that the transit-timing variation of XO-3b is due to tidal orbital decay, is $(1.50 \pm 0.03)\times 10^5$ \citep{yang2021} --- statistically identical to the measurement we obtained for WASP-12.

Orbital decay has not yet been detected for any other planet, at least not with nearly the same high confidence level as it has been detected for WASP-12b. However, looking at the wider population of hot Jupiters, there are several other planets with characteristics similar to those of WASP-12b that might be expected to show comparable rates of orbital decay. In particular, if we assume Equation~\eqref{eq:dpdt} is correct and that $Q'_*$ is the same for all hot-Jupiter systems, then WASP-103b, KELT-16b, WASP-18b, and the recently discovered ultra-hot Jupiter TOI-2109b have larger predicted $|dP/dt|$ values than WASP-12b (see Figure~14 in \citealt{wong2021toi2109}). Notably, the hosts of all five aforementioned planets are F-type stars. The transits of WASP-18b have been monitored for over 15 years, with no sign yet of a decaying orbital period; this nondetection constrains WASP-18's tidal quality factor to $Q'_* > 3.9 \times 10^6$ at $2\sigma$ \citep{m20a}. Meanwhile, a recent analysis of WASP-103b transit timings spanning seven years yielded a $3\sigma$ lower limit of $Q'_* > 1.6 \times 10^6$ \citep{barros2022}. Perhaps there is a diversity in the dissipative properties of planet-hosting stars, or a very sensitive dependence on the forcing frequency and internal structure of the star.

TOI-2109b is a particularly promising candidate for probing tidal orbital decay. Due to its ultra-short 0.67 day orbital period and its large mass of roughly $5\,M_{\rm Jup}$, this extreme hot Jupiter has a predicted orbital decay rate that is more than an order of magnitude faster than that of WASP-12b (assuming the same $Q'_*$). Follow-up transit observations of the TOI-2109 system over the coming years should soon allow us to determine if the host star has a $Q'_*$ value comparable to that of WASP-12, or a higher $Q'_*$ similar to WASP-18. 

\begin{figure*}[t]
\centering
\includegraphics[width=\linewidth]{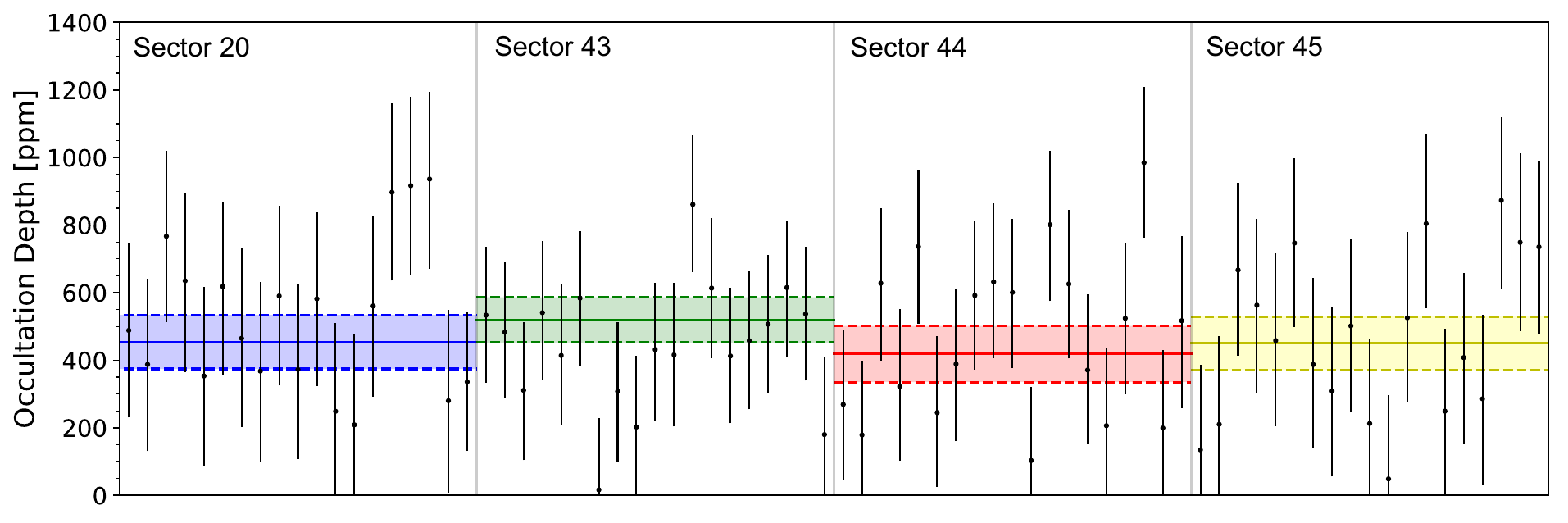}
\caption{Plot of the individual secondary eclipse depths measured from the TESS light curve (black). The datapoints are arranged sequentially by epoch, with occultations that coincided with momentum dumps removed. The colored lines and shaded areas indicate the median eclipse depths and $1\sigma$ confidence regions that we derived from full-orbit phase-curve fits for each TESS sector (see Table~\ref{tab:sectors}). The scatter in the individual eclipse depths is comparable to the typical uncertainties, and the depths measured for each sector are mutually consistent at better than the $1\sigma$ level.}
\label{fig:ecls}
\end{figure*}

\subsection{Constraints on Atmospheric Variability}\label{subsec:variability}

As was discussed in the Introduction, several prior secondary eclipse observations of WASP-12b have yielded discrepant depth measurements, raising the possibility of time-varying behavior of the planet's atmosphere.  However, most of these previous results were obtained using different ground-based facilities, and as such it is difficult to ascertain the extent to which variable observing conditions and photometric extraction methodologies may have caused the formal uncertainties in the occultation depths to be underestimated.

\citet{wong2021year2} fit the individual WASP-12b secondary eclipses from the TESS Sector 20 light curve and found that the measured depths show no sign of variability to within the measurement precision. We expanded this analysis to all four TESS sectors. After dividing out the best-fit systematics model from each sector's light curve, we fixed the mid-eclipse times based on the best-fit decaying transit ephemeris from Equation~\eqref{eqn:ephem}. The parameters $b$, $a/R_*$, and $R_p/R_*$ were fixed to the best-fit values computed from our combined phase-curve fit (Table~\ref{tab:sectors}), and the per-point photometric uncertainties were set to the respective red-noise-enhanced $\sigma_r$ values. No additional systematics modeling was included.

Figure~\ref{fig:ecls} shows the depths measured from the 76 secondary eclipse light curves, arranged sequentially by sector. The occultation depths derived from the individual sector phase-curve fits are also shown. Due to the low signal-to-noise ratio of the eclipse light curves, the precision of the individual depth measurements is poor. Nevertheless, we can still place broad constraints on any orbit-to-orbit atmospheric variability that may be present. The majority of individual eclipse depths agree with the corresponding sector-wide values at better than the $1\sigma$ level. The largest discrepancy between a pair of eclipse depths is $3.1\sigma$. The overall scatter in the measurements (223 ppm) is smaller than the median uncertainty (242 ppm), indicating that the individual eclipse depths are consistent with being drawn from the same distribution. We calculated the Lomb--Scargle periodogram for the eclipse measurements from the three consecutive sectors (43--45) and detected no significant periodicities. We are able to place a $2\sigma$ upper limit of roughly 450 ppm on the orbit-to-orbit variability of WASP-12's dayside brightness in the TESS bandpass. 

We now compare this constraint to the previously reported instances of variable secondary eclipse depths. The most significant discrepancy in the literature lies between two $V$-band observations in \citet{vonessen2019}: $280 \pm 100$ and $1160 \pm 170$ ppm. This $4.5\sigma$ difference greatly exceeds our upper limit; moreover, given the lower planet--star contrast ratio in the $V$-band, such a discrepancy in secondary eclipse depth would translate to an even larger absolute change in the TESS-band dayside brightness. It is important to mention that a third $V$-band eclipse light curve from the same instrument yielded a nondetection ($-30 \pm 170$ ppm). Therefore, barring the serendipitous observation of a hitherto unexplained astrophysical event, we suggest that uncorrected instrumental systematics and/or observing conditions led to significant biases in one or both of the reported $V$-band secondary eclipse depths in \citet{vonessen2019}. 

The remaining instances of short-wavelength ($<$1 $\mu$m) secondary eclipse depth discrepancies in the literature are smaller in magnitude. \citet{hooton2019} measured $i'$-band depths of $440 \pm 210$ and $970 \pm 140$ mmag using two different telescopes. The former value is in good agreement with our TESS-band sector-wide depth measurements. The $2.1\sigma$ difference between the two depths is only marginally larger than our $2\sigma$ upper limit. Lastly, the three $z'$-band eclipse depths published in \citet{lopez2010} and \citet{fohring2013} are $680 \pm 210$, $1000 \pm 230$, and $1300 \pm 130$ ppm, with a $2.5\sigma$ mutual discrepancy between the largest and smallest values. In light of the relatively low statistical significance of these mismatches, as well as the lack of uniformity in instrumentation and possible $i'$- and $z'$-band telluric contamination in ground-based observations, we should be cautious about interpreting these measurements as evidence for atmospheric variability.

The TESS light curves provide much tighter constraints on longer-term variability of WASP-12b's atmosphere. The phase-curve results that we obtained for the individual sectors offer four independent snapshots of the planet's properties, each averaged over a roughly one-month period. As illustrated in Figure~\ref{fig:ecls}, the best-fit secondary eclipse depths for the four sectors are in excellent agreement, with a standard deviation of 37 ppm and a median uncertainty of 80 ppm. Moreover, there is no evidence of a systematic shift in the size of the occultation across the roughly 600 days that elapsed between Sector 20 and Sectors 43--45. We place a $2\sigma$ upper limit of $\approx$80 ppm on the TESS-band dayside atmospheric variability of WASP-12b across both month-long and year-long timescales. In other words, the planet's dayside brightness is invariable to within $\approx$17\%. Likewise, there are no significant variations in the day--night temperature contrast, with the atmospheric brightness amplitudes ($A_{\rm atm}$) from the four TESS sectors consistent to within $2.2\sigma$ (see Table \ref{tab:sectors}).

Our Palomar/WIRC $K_s$-band occultation measurement ($2810 \pm 390$ ppm; Section~\ref{subsec:eclipses}) adds to the appreciable body of previous $\sim$2.2 $\mu$m observations. The first $K_s$-band eclipse depth was published in \citet{croll2011}: $3090 \pm 130$ ppm. \citet{zhao2012} obtained two secondary eclipse measurements in the $K_s$ band with the TIFKAM instrument on the Hiltner Telescope: $2810 \pm 850$ and $3160 \pm 950$ ppm, with a joint depth of $2990 \pm 640$ ppm. Narrowband 2.312 $\mu$m observations at the Subaru Observatory yielded a notably deeper occultation depth of $4500 \pm 600$ ppm \citep{crossfield2012}. \citet{croll2015} used the Canada--France--Hawaii Telescope/Wide-field InfraRed Camera (CFHT/WIRFCam) instrument on three nights and measured $K_s$-band secondary eclipse depths of $2840 \pm 200$, $2890 \pm 180$, and $2590 \pm 420$ ppm; the same team also observed the system in the $K_{\rm CONT}$ band ($\sim$2.22 $\mu$m) and calculated a depth of $2640 \pm 450$ ppm.  The new $K_s$-band measurement is consistent with all prior values, except for the significantly larger depth reported by \citet{crossfield2012}. 

\section{Summary}\label{sec:conclusion}

In this paper, we analyzed the full TESS light curve of WASP-12, including three sectors' worth of photometry obtained during the spacecraft's ongoing extended mission. We also presented a new $K_s$-band secondary eclipse observation taken with the Palomar/WIRC instrument. Below, we reprise the main results of our work.
\begin{enumerate}
\vspace{-0.2cm}
\item From our combined fit to all four sectors of TESS data (Section~\ref{subsec:phase}; Table~\ref{tab:sectors}), we significantly improved the precision of the measured phase-curve properties. We obtained a secondary eclipse depth of $466 \pm 35$ ppm, a peak-to-peak atmospheric brightness amplitude of $524 \pm 30$ ppm, and a nightside flux that is consistent with zero. A marginal phase-curve offset of $6\overset{\circ}{.}2 \pm 2\overset{\circ}{.}8$ was also detected. The observed ellipsoidal distortion amplitude of $65 \pm 14$ ppm is consistent with theoretical predictions based on the measured planet--star mass ratio. \vspace{-0.2cm}
\item We fit for 74 individual transit timings (Section~\ref{subsec:transits}; Table~\ref{tab:transits}) and five mid-occultation times (four from the TESS light curves and one from the Palomar/WIRC observation; Section~\ref{subsec:eclipses}). The updated orbital period is $1.091419370 \pm 0.000000020$ days, with a decay rate of $-29.81 \pm 0.94$ ms yr$^{-1}$.
\item From the sector-wide phase-curve fits, we found close agreement in the measured secondary eclipse depths, planetary phase-curve amplitudes, and phase-curve offsets. We placed a tight $2\sigma$ upper limit of $\approx$80 ppm ($\approx$17\%) on the dayside atmospheric brightness variability in the TESS bandpass occurring on month-long and year-long timescales.
\item The set of individual TESS-band secondary eclipse depths shows no signs of variability (Figure~\ref{fig:ecls}). We placed a broad $2\sigma$ upper limit of 450 ppm on the orbit-to-orbit variability of WASP-12b's dayside brightness. 
\item Our new $K_s$-band occultation depth measurement ($2810 \pm 390$ ppm) agrees with all previously published values in the literature, except for the anomalously high value from \citet{crossfield2012}, which we consider to be an outlier.
\end{enumerate}

%=============================================================================

Funding for the TESS mission is provided by NASA’s Science Mission directorate. This paper includes data collected by the TESS mission, which are publicly available from the Mikulski Archive for Space Telescopes. Resources supporting this work were provided by the NASA High-End Computing (HEC) Program through the NASA Advanced Supercomputing (NAS) Division at Ames Research Center for the production of the SPOC data products. We thank the Palomar Observatory team, particularly Paul~Nied, Carolyn~Heffner, and Tom~Barlow, for enabling the $K_s$-band secondary eclipse observations and facilitating remote operations on the Hale $200''$ Telescope. I.W. is supported by an appointment to the NASA Postdoctoral Program at the NASA Goddard Space Flight Center, administered by the Universities Space Research Association under contract with NASA. S.V. is supported by an NSF Graduate Research Fellowship. H.A.K. acknowledges support from NSF CAREER grant 1555095.

\facilities{TESS, Palomar/WIRC.}

%===============================================================================
\software{
\texttt{batman} \citep{batman},
\texttt{emcee} \citep{emcee},
\texttt{exoplanet} \citep{exoplanet},
\texttt{ExoTEP} \citep{benneke2019,wong2020hatp12}.}

%===============================================================================

\appendix 

\section{Results from Individual TESS Transit Fits}\label{sec:transitres}
\restartappendixnumbering

\begin{figure*}[b]
\centering
\includegraphics[width=0.7\linewidth]{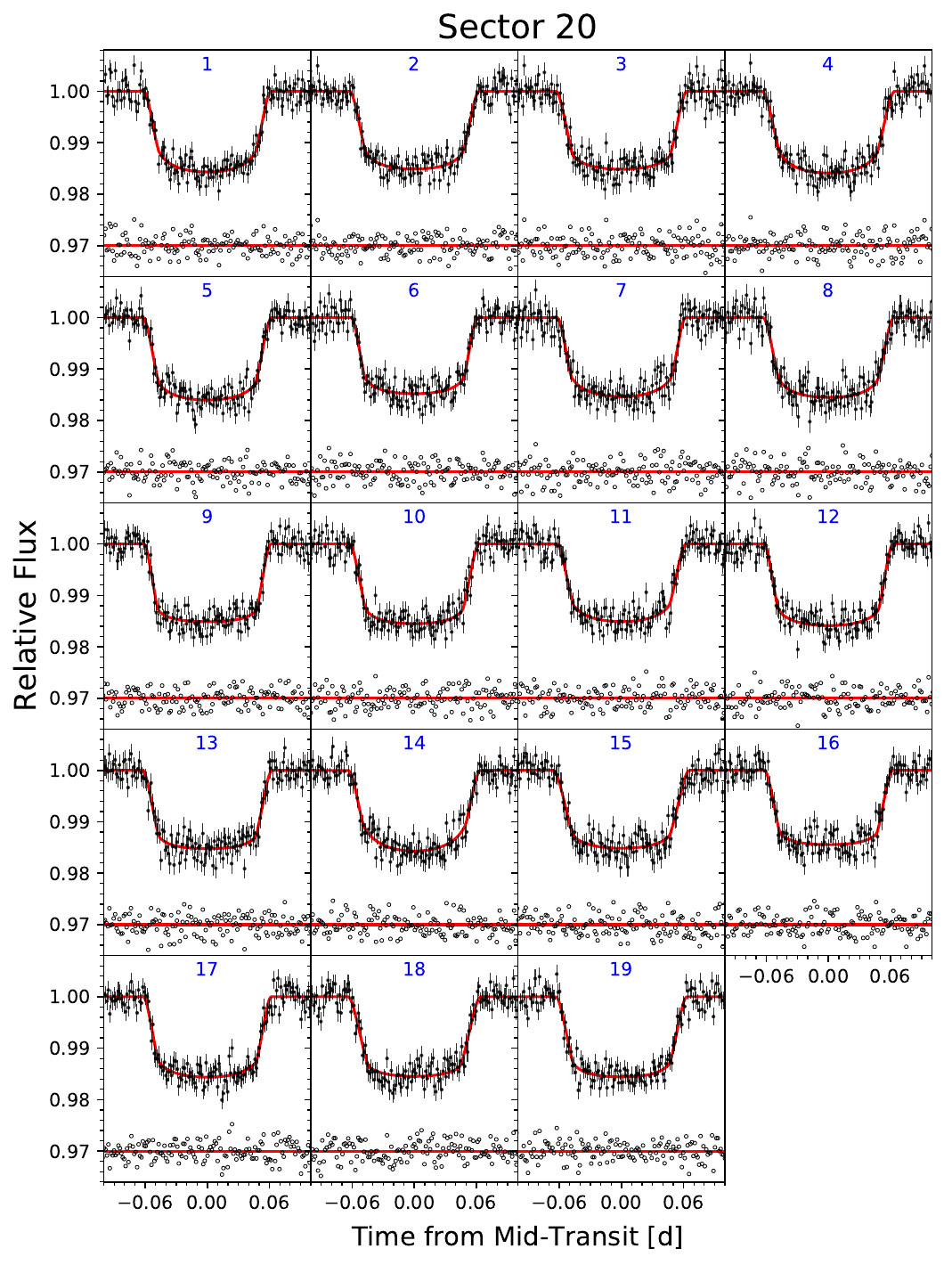}
\caption{Compilation of the TESS Sector 20 transit light curves. The systematics-corrected photometry is shown in black. The red curves are the best-fit models from our individual transit fits. The corresponding residuals from the best-fit models are shown (offset by 0.03) in each panel.}
\label{fig:trans20}
\end{figure*}

\begin{figure*}[t]
\centering
\includegraphics[width=0.9\linewidth]{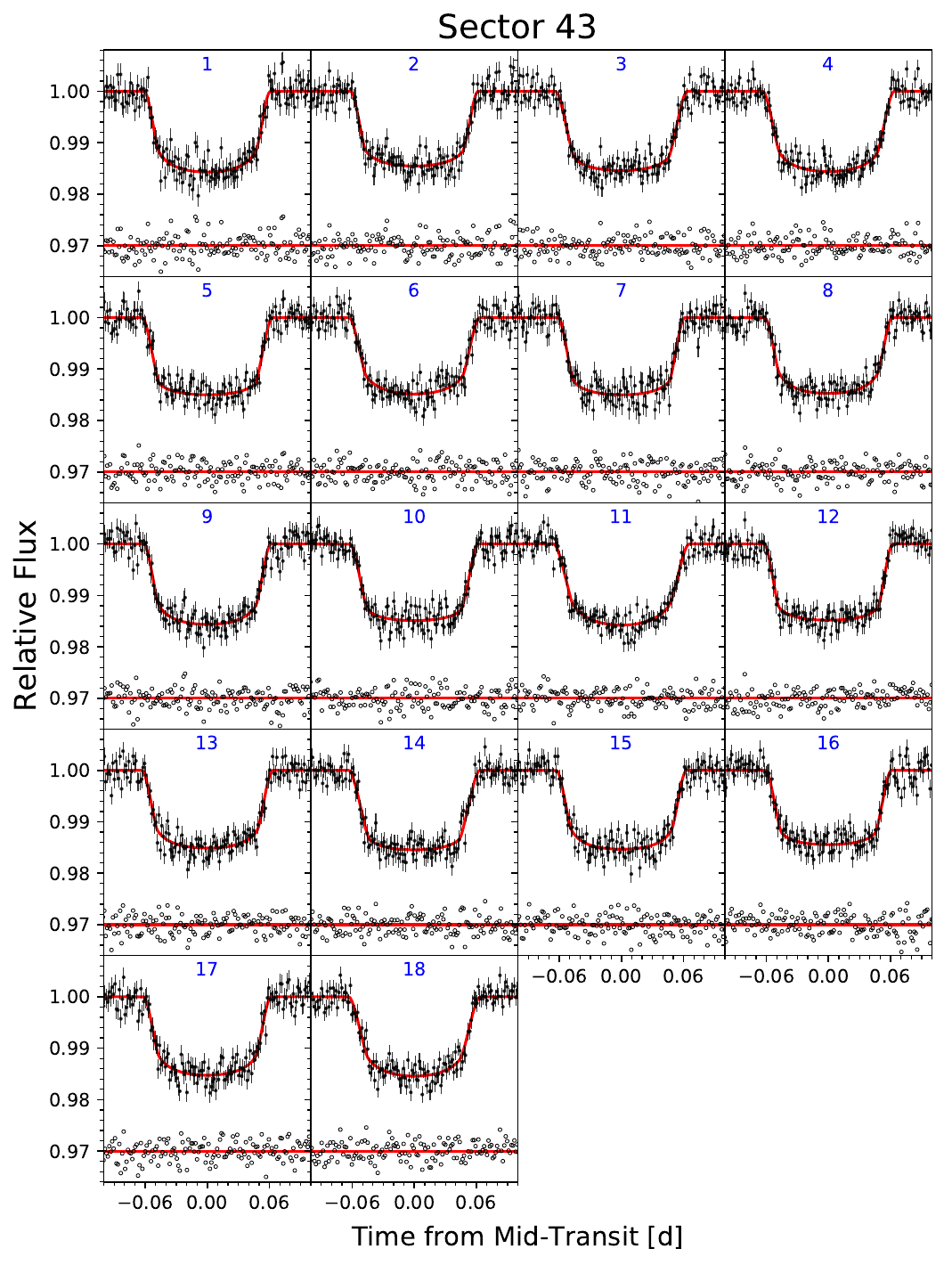}
\caption{Same as Figure~\ref{fig:trans20}, but for the Sector 43 transits.}
\label{fig:trans43}
\end{figure*}

\begin{figure*}[t]
\centering
\includegraphics[width=0.9\linewidth]{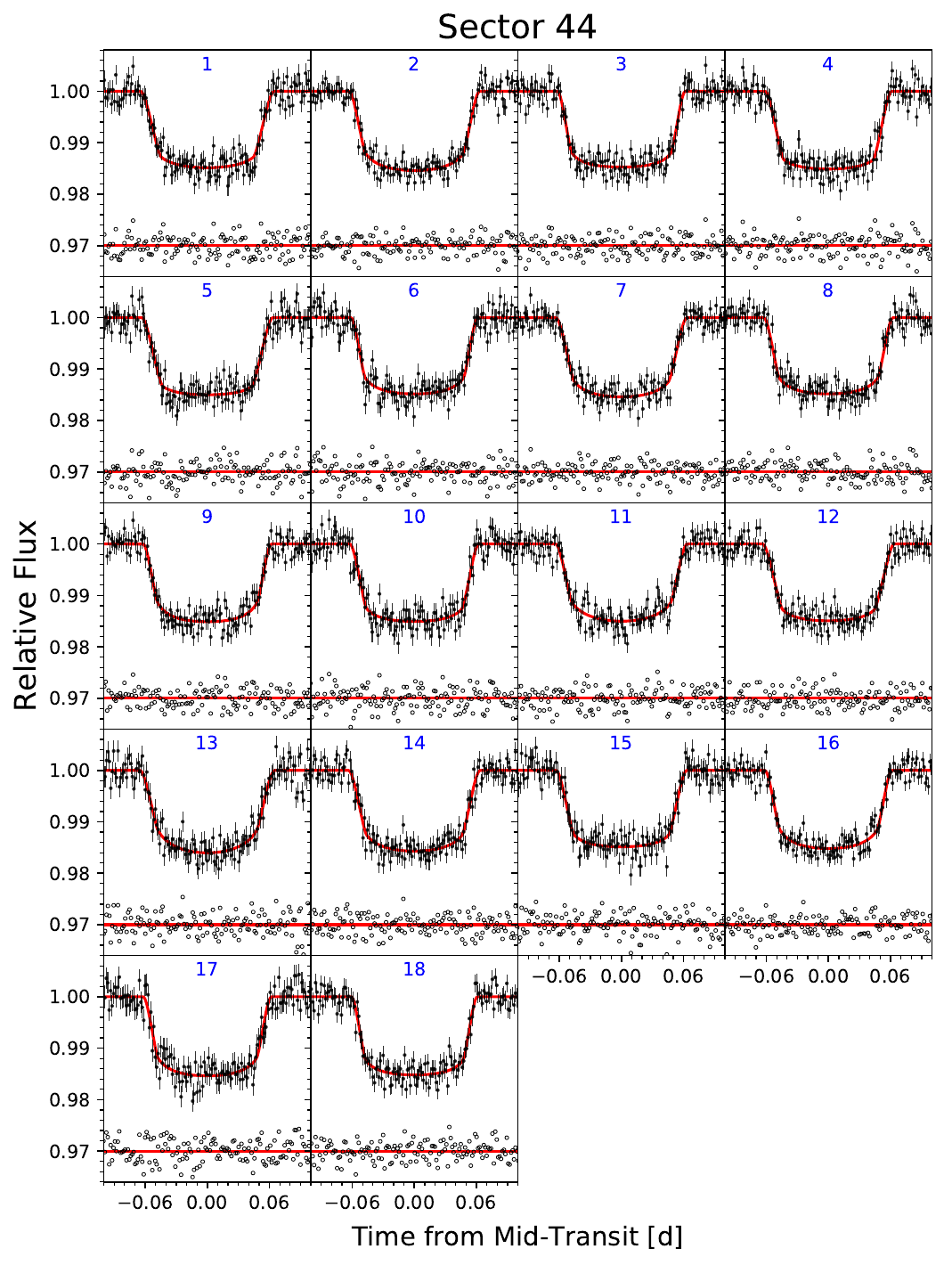}
\caption{Same as Figure~\ref{fig:trans20}, but for the Sector 44 transits.}
\label{fig:trans44}
\end{figure*}

\begin{figure*}[t]
\centering
\includegraphics[width=0.9\linewidth]{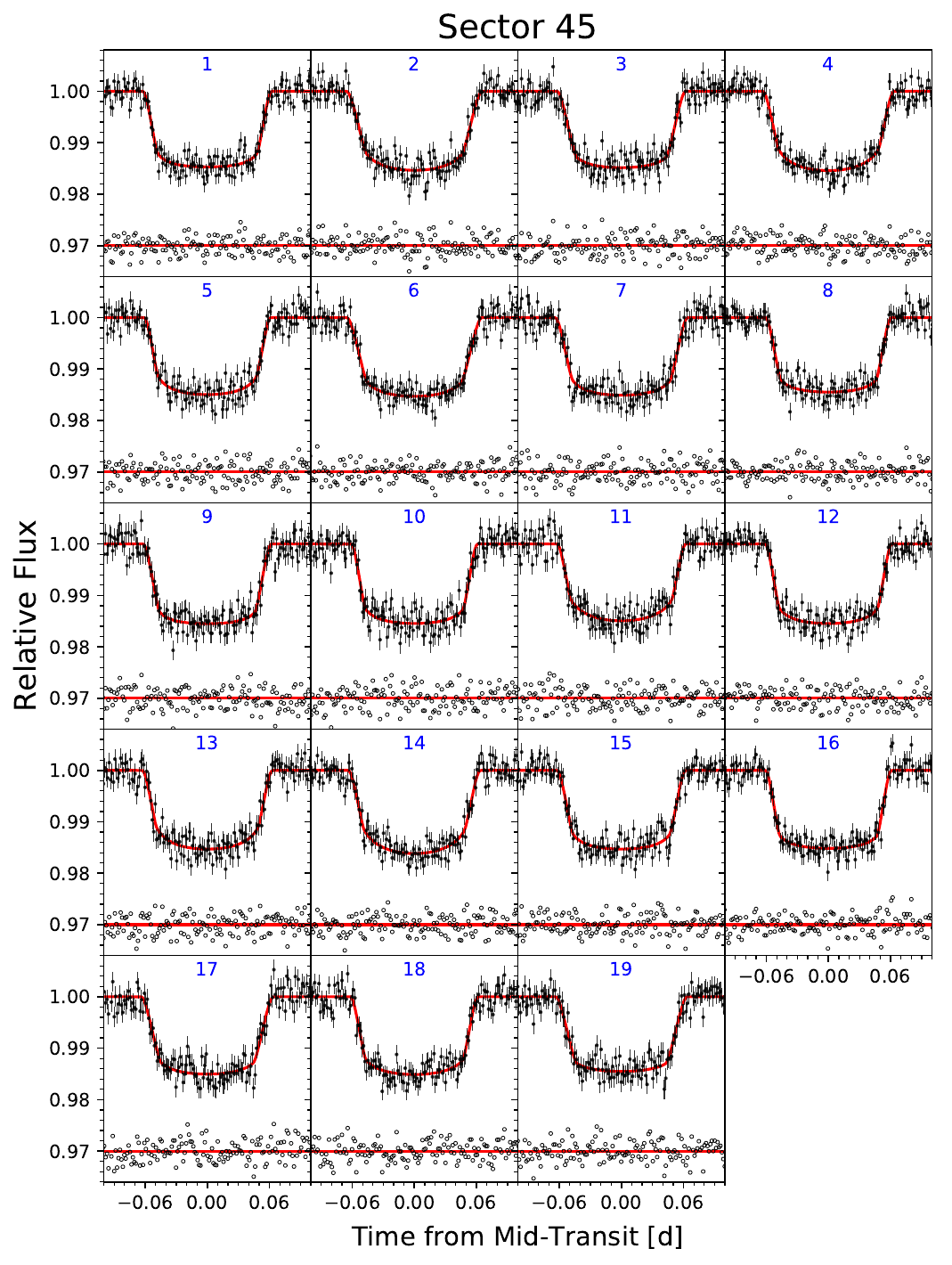}
\caption{Same as Figure~\ref{fig:trans20}, but for the Sector 45 transits.}
\label{fig:trans45}
\end{figure*}

The transit light curves and best-fit models are plotted in Figures~\ref{fig:trans20}--\ref{fig:trans45}. Table~\ref{tab:transits} lists the results from our individual TESS transit light-curve fits. Each transit is labeled by the sector and sequential transit number; Figure~\ref{fig:raw} shows the location of the corresponding transit events. The second through fifth columns provide the median and $1\sigma$ values for the mid-transit time $T_c$, planet--star radius ratio $R_{p}/R_{*}$, impact parameter $b$, and scaled orbital semimajor axis $a/R_{*}$. The fitted transit depths and transit-shape parameter values are mutually consistent to within 2.3--2.5$\sigma$. 

\clearpage

\startlongtable
\begin{deluxetable*}{ccccc}
\tablewidth{0pc}
\setlength{\tabcolsep}{14pt}
\renewcommand{\arraystretch}{0.95}
\tabletypesize{\footnotesize}
\tablecaption{
    Results of Individual Transit Fits
    \label{tab:transits}
}
\tablehead{
    \colhead{Transit Number}                     &
    \colhead{$T_c$ (${\rm BJD}_{\rm TDB}$)}                     &
    \colhead{$R_p/R_*$} &
    \colhead{$b$}  &
    \colhead{$a/R_*$} 
}
\startdata
S20-1 & $2{,}458{,}844.09668 \pm 0.00046$ & $0.1168 \pm 0.0024$ & $0.26 \pm 0.19$ & $3.17 \pm 0.20$ \\
S20-2 & $2{,}458{,}845.18772 \pm 0.00042$ & $0.1152 \pm 0.0022$ & $0.31 \pm 0.20$ & $3.10 \pm 0.19$ \\
S20-3 & $2{,}458{,}846.27969 \pm 0.00053$ & $0.1162 \pm 0.0024$ & $0.30 \pm 0.21$ & $3.07 \pm 0.22$ \\
S20-4 & $2{,}458{,}847.37086 \pm 0.00049$ & $0.1181 \pm 0.0022$ & $0.32 \pm 0.20$ & $3.08 \pm 0.27$ \\
S20-5 & $2{,}458{,}848.46246 \pm 0.00047$ & $0.1188 \pm 0.0019$ & $0.23 \pm 0.18$ & $3.16 \pm 0.19$ \\
S20-6 & $2{,}458{,}849.55305 \pm 0.00045$ & $0.1143 \pm 0.0016$ & $0.17 \pm 0.18$ & $3.20 \pm 0.14$ \\
S20-7 & $2{,}458{,}850.64500 \pm 0.00054$ & $0.1153 \pm 0.0020$ & $0.21 \pm 0.18$ & $3.15 \pm 0.16$ \\
S20-8 & $2{,}458{,}851.73597 \pm 0.00054$ & $0.1177 \pm 0.0025$ & $0.33 \pm 0.22$ & $3.04 \pm 0.26$ \\
S20-9 & $2{,}458{,}852.82816 \pm 0.00044$ & $0.1174 \pm 0.0014$ & $0.13 \pm 0.13$ & $3.20 \pm 0.07$ \\
S20-10 & $2{,}458{,}853.91929 \pm 0.00045$ & $0.1179 \pm 0.0020$ & $0.30 \pm 0.19$ & $3.10 \pm 0.21$ \\
S20-11 & $2{,}458{,}859.37709 \pm 0.00049$ & $0.1155 \pm 0.0020$ & $0.31 \pm 0.19$ & $3.11 \pm 0.20$ \\
S20-12 & $2{,}458{,}860.46813 \pm 0.00045$ & $0.1182 \pm 0.0016$ & $0.19 \pm 0.15$ & $3.14 \pm 0.11$ \\
S20-13 & $2{,}458{,}861.55807 \pm 0.00047$ & $0.1168 \pm 0.0018$ & $0.28 \pm 0.17$ & $3.11 \pm 0.17$ \\
S20-14 & $2{,}458{,}862.65051 \pm 0.00055$ & $0.1161 \pm 0.0023$ & $0.27 \pm 0.18$ & $3.04 \pm 0.18$ \\
S20-15 & $2{,}458{,}863.74202 \pm 0.00058$ & $0.1169 \pm 0.0027$ & $0.45 \pm 0.29$ & $2.86 \pm 0.25$ \\
S20-16 & $2{,}458{,}864.83386 \pm 0.00049$ & $0.1145 \pm 0.0018$ & $0.21 \pm 0.17$ & $3.18 \pm 0.15$ \\
S20-17 & $2{,}458{,}865.92430 \pm 0.00047$ & $0.1174 \pm 0.0019$ & $0.23 \pm 0.16$ & $3.16 \pm 0.16$ \\
S20-18 & $2{,}458{,}867.01609 \pm 0.00051$ & $0.1174 \pm 0.0020$ & $0.33 \pm 0.22$ & $3.06 \pm 0.22$ \\
S20-19 & $2{,}458{,}868.10758 \pm 0.00043$ & $0.1178 \pm 0.0018$ & $0.29 \pm 0.21$ & $3.08 \pm 0.18$ \\
S43-1 & $2{,}459{,}474.93462 \pm 0.00046$ & $0.1166 \pm 0.0019$ & $0.22 \pm 0.16$ & $3.20 \pm 0.14$ \\
S43-2 & $2{,}459{,}476.02746 \pm 0.00056$ & $0.1135 \pm 0.0019$ & $0.25 \pm 0.18$ & $3.15 \pm 0.19$ \\
S43-3 & $2{,}459{,}477.11831 \pm 0.00050$ & $0.1170 \pm 0.0024$ & $0.38 \pm 0.23$ & $3.00 \pm 0.21$ \\
S43-4 & $2{,}459{,}479.30082 \pm 0.00047$ & $0.1168 \pm 0.0020$ & $0.35 \pm 0.17$ & $3.02 \pm 0.17$ \\
S43-5 & $2{,}459{,}480.39281 \pm 0.00048$ & $0.1153 \pm 0.0018$ & $0.28 \pm 0.17$ & $3.09 \pm 0.14$ \\
S43-6 & $2{,}459{,}481.48348 \pm 0.00051$ & $0.1145 \pm 0.0021$ & $0.32 \pm 0.21$ & $3.09 \pm 0.21$ \\
S43-7 & $2{,}459{,}482.57475 \pm 0.00049$ & $0.1153 \pm 0.0018$ & $0.21 \pm 0.17$ & $3.19 \pm 0.15$ \\
S43-8 & $2{,}459{,}483.66690 \pm 0.00051$ & $0.1135 \pm 0.0020$ & $0.27 \pm 0.18$ & $3.15 \pm 0.19$ \\
S43-9 & $2{,}459{,}484.75830 \pm 0.00048$ & $0.1167 \pm 0.0020$ & $0.28 \pm 0.16$ & $3.14 \pm 0.19$ \\
S43-10 & $2{,}459{,}488.03344 \pm 0.00046$ & $0.1155 \pm 0.0016$ & $0.24 \pm 0.17$ & $3.17 \pm 0.16$ \\
S43-11 & $2{,}459{,}489.12326 \pm 0.00051$ & $0.1188 \pm 0.0031$ & $0.51 \pm 0.25$ & $2.76 \pm 0.29$ \\
S43-12 & $2{,}459{,}490.21514 \pm 0.00042$ & $0.1146 \pm 0.0015$ & $0.17 \pm 0.14$ & $3.15 \pm 0.09$ \\
S43-13 & $2{,}459{,}491.30624 \pm 0.00048$ & $0.1145 \pm 0.0016$ & $0.20 \pm 0.13$ & $3.12 \pm 0.10$ \\
S43-14 & $2{,}459{,}492.39827 \pm 0.00051$ & $0.1170 \pm 0.0022$ & $0.28 \pm 0.20$ & $3.14 \pm 0.22$ \\
S43-15 & $2{,}459{,}493.48919 \pm 0.00043$ & $0.1163 \pm 0.0017$ & $0.25 \pm 0.17$ & $3.10 \pm 0.16$ \\
S43-16 & $2{,}459{,}494.58116 \pm 0.00052$ & $0.1134 \pm 0.0020$ & $0.28 \pm 0.20$ & $3.17 \pm 0.21$ \\
S43-17 & $2{,}459{,}495.67260 \pm 0.00047$ & $0.1156 \pm 0.0021$ & $0.26 \pm 0.20$ & $3.16 \pm 0.23$ \\
S43-18 & $2{,}459{,}496.76348 \pm 0.00049$ & $0.1165 \pm 0.0026$ & $0.46 \pm 0.20$ & $2.94 \pm 0.21$ \\
S44-1 & $2{,}459{,}501.12937 \pm 0.00051$ & $0.1164 \pm 0.0024$ & $0.45 \pm 0.19$ & $2.89 \pm 0.21$ \\
S44-2 & $2{,}459{,}502.22115 \pm 0.00043$ & $0.1161 \pm 0.0019$ & $0.27 \pm 0.18$ & $3.10 \pm 0.17$ \\
S44-3 & $2{,}459{,}503.31271 \pm 0.00045$ & $0.1149 \pm 0.0014$ & $0.16 \pm 0.15$ & $3.16 \pm 0.10$ \\
S44-4 & $2{,}459{,}504.40350 \pm 0.00047$ & $0.1158 \pm 0.0022$ & $0.25 \pm 0.21$ & $3.23 \pm 0.23$ \\
S44-5 & $2{,}459{,}505.49468 \pm 0.00059$ & $0.1168 \pm 0.0025$ & $0.48 \pm 0.21$ & $2.90 \pm 0.23$ \\
S44-6 & $2{,}459{,}506.58620 \pm 0.00051$ & $0.1143 \pm 0.0018$ & $0.23 \pm 0.16$ & $3.16 \pm 0.16$ \\
S44-7 & $2{,}459{,}507.67798 \pm 0.00045$ & $0.1165 \pm 0.0020$ & $0.42 \pm 0.15$ & $2.98 \pm 0.16$ \\
S44-8 & $2{,}459{,}508.76993 \pm 0.00048$ & $0.1140 \pm 0.0016$ & $0.21 \pm 0.13$ & $3.15 \pm 0.09$ \\
S44-9 & $2{,}459{,}509.86066 \pm 0.00045$ & $0.1154 \pm 0.0018$ & $0.23 \pm 0.17$ & $3.15 \pm 0.15$ \\
S44-10 & $2{,}459{,}510.95217 \pm 0.00056$ & $0.1162 \pm 0.0022$ & $0.31 \pm 0.15$ & $3.09 \pm 0.19$ \\
S44-11 & $2{,}459{,}515.31724 \pm 0.00049$ & $0.1144 \pm 0.0024$ & $0.35 \pm 0.20$ & $3.02 \pm 0.25$ \\
S44-12 & $2{,}459{,}516.40891 \pm 0.00053$ & $0.1150 \pm 0.0022$ & $0.35 \pm 0.20$ & $3.06 \pm 0.24$ \\
S44-13 & $2{,}459{,}517.50098 \pm 0.00060$ & $0.1191 \pm 0.0025$ & $0.48 \pm 0.16$ & $2.83 \pm 0.20$ \\
S44-14 & $2{,}459{,}518.59247 \pm 0.00048$ & $0.1181 \pm 0.0021$ & $0.37 \pm 0.21$ & $3.01 \pm 0.19$ \\
S44-15 & $2{,}459{,}519.68224 \pm 0.00057$ & $0.1151 \pm 0.0022$ & $0.34 \pm 0.20$ & $3.07 \pm 0.19$ \\
S44-16 & $2{,}459{,}521.86571 \pm 0.00042$ & $0.1153 \pm 0.0018$ & $0.25 \pm 0.17$ & $3.16 \pm 0.17$ \\
S44-17 & $2{,}459{,}522.95725 \pm 0.00057$ & $0.1168 \pm 0.0021$ & $0.27 \pm 0.19$ & $3.12 \pm 0.23$ \\
S44-18 & $2{,}459{,}524.04947 \pm 0.00039$ & $0.1162 \pm 0.0016$ & $0.18 \pm 0.16$ & $3.20 \pm 0.12$ \\
S45-1 & $2{,}459{,}527.32351 \pm 0.00043$ & $0.1151 \pm 0.0016$ & $0.24 \pm 0.15$ & $3.11 \pm 0.15$ \\
S45-2 & $2{,}459{,}528.41517 \pm 0.00054$ & $0.1179 \pm 0.0024$ & $0.52 \pm 0.13$ & $2.75 \pm 0.19$ \\
S45-3 & $2{,}459{,}529.50671 \pm 0.00055$ & $0.1156 \pm 0.0022$ & $0.44 \pm 0.21$ & $2.99 \pm 0.22$ \\
S45-4 & $2{,}459{,}530.59839 \pm 0.00057$ & $0.1167 \pm 0.0030$ & $0.43 \pm 0.28$ & $2.92 \pm 0.25$ \\
S45-5 & $2{,}459{,}531.68987 \pm 0.00052$ & $0.1145 \pm 0.0016$ & $0.22 \pm 0.15$ & $3.19 \pm 0.14$ \\
S45-6 & $2{,}459{,}532.78034 \pm 0.00048$ & $0.1170 \pm 0.0024$ & $0.45 \pm 0.19$ & $2.87 \pm 0.20$ \\
S45-7 & $2{,}459{,}533.87187 \pm 0.00054$ & $0.1157 \pm 0.0019$ & $0.32 \pm 0.23$ & $3.03 \pm 0.20$ \\
S45-8 & $2{,}459{,}534.96448 \pm 0.00046$ & $0.1135 \pm 0.0015$ & $0.21 \pm 0.16$ & $3.21 \pm 0.14$ \\
S45-9 & $2{,}459{,}536.05478 \pm 0.00052$ & $0.1179 \pm 0.0018$ & $0.27 \pm 0.16$ & $3.14 \pm 0.18$ \\
S45-10 & $2{,}459{,}537.14683 \pm 0.00055$ & $0.1165 \pm 0.0016$ & $0.21 \pm 0.16$ & $3.18 \pm 0.14$ \\
S45-11 & $2{,}459{,}540.42054 \pm 0.00047$ & $0.1146 \pm 0.0017$ & $0.17 \pm 0.15$ & $3.17 \pm 0.11$ \\
S45-12 & $2{,}459{,}541.51151 \pm 0.00041$ & $0.1168 \pm 0.0015$ & $0.15 \pm 0.12$ & $3.20 \pm 0.08$ \\
S45-13 & $2{,}459{,}542.60447 \pm 0.00048$ & $0.1155 \pm 0.0017$ & $0.22 \pm 0.17$ & $3.11 \pm 0.14$ \\
S45-14 & $2{,}459{,}543.69504 \pm 0.00046$ & $0.1171 \pm 0.0020$ & $0.28 \pm 0.18$ & $3.06 \pm 0.16$ \\
S45-15 & $2{,}459{,}545.87708 \pm 0.00049$ & $0.1164 \pm 0.0021$ & $0.28 \pm 0.20$ & $3.12 \pm 0.21$ \\
S45-16 & $2{,}459{,}546.96868 \pm 0.00044$ & $0.1158 \pm 0.0015$ & $0.12 \pm 0.12$ & $3.24 \pm 0.06$ \\
S45-17 & $2{,}459{,}548.06099 \pm 0.00055$ & $0.1167 \pm 0.0022$ & $0.46 \pm 0.18$ & $2.90 \pm 0.20$ \\
S45-18 & $2{,}459{,}549.15196 \pm 0.00049$ & $0.1160 \pm 0.0017$ & $0.18 \pm 0.16$ & $3.19 \pm 0.13$ \\
S45-19 & $2{,}459{,}550.24370 \pm 0.00054$ & $0.1143 \pm 0.0024$ & $0.41 \pm 0.23$ & $2.98 \pm 0.25$ \\
\enddata
\end{deluxetable*}

\end{document}